
%
\expandafter\ifx\csname phyzzx\endcsname\relax\else
 \errhelp{Hit <CR> and go ahead.}
 \errmessage{PHYZZX macros are already loaded or input. }
 \endinput \fi
\catcode`\@=11 
%
%
%
\font\seventeenrm=cmr17
\font\fourteenrm=cmr12 scaled\magstep1
\font\twelverm=cmr12
\font\ninerm=cmr9            \font\sixrm=cmr6
%
\font\fourteenbf=cmbx10 scaled\magstep2
\font\twelvebf=cmbx12
\font\ninebf=cmbx9            \font\sixbf=cmbx6
%
\font\fourteeni=cmmi10 scaled\magstep2      \skewchar\fourteeni='177
\font\twelvei=cmmi12			        \skewchar\twelvei='177
\font\ninei=cmmi9                           \skewchar\ninei='177
\font\sixi=cmmi6                            \skewchar\sixi='177
%
\font\fourteensy=cmsy10 scaled\magstep2     \skewchar\fourteensy='60
\font\twelvesy=cmsy10 scaled\magstep1	    \skewchar\twelvesy='60
\font\ninesy=cmsy9                          \skewchar\ninesy='60
\font\sixsy=cmsy6                           \skewchar\sixsy='60
%
\font\fourteenex=cmex10 scaled\magstep2
\font\twelveex=cmex10 scaled\magstep1
%
\font\fourteensl=cmsl12 scaled\magstep1
\font\twelvesl=cmsl12
\font\ninesl=cmsl9
%
\font\fourteenit=cmti12 scaled\magstep1
\font\twelveit=cmti12
\font\nineit=cmti9
\font\fourteentt=cmtt10 scaled\magstep2
\font\twelvett=cmtt12
\font\fourteencp=cmcsc10 scaled\magstep2
\font\twelvecp=cmcsc10 scaled\magstep1
\font\tencp=cmcsc10
\newfam\cpfam
\newdimen\b@gheight		\b@gheight=12pt
\newcount\f@ntkey		\f@ntkey=0
\def\f@m{\afterassignment\samef@nt\f@ntkey=}
\def\samef@nt{\fam=\f@ntkey \the\textfont\f@ntkey\relax}
\def\rm{\f@m0 }
\def\mit{\f@m1 }         
\def\cal{\f@m2 }
\def\it{\f@m\itfam}
\def\sl{\f@m\slfam}
\def\bf{\f@m\bffam}
\def\tt{\f@m\ttfam}
\def\caps{\f@m\cpfam}
\def\fourteenpoint{\relax
    \textfont0=\fourteenrm          \scriptfont0=\tenrm
      \scriptscriptfont0=\sevenrm
    \textfont1=\fourteeni           \scriptfont1=\teni
      \scriptscriptfont1=\seveni
    \textfont2=\fourteensy          \scriptfont2=\tensy
      \scriptscriptfont2=\sevensy
    \textfont3=\fourteenex          \scriptfont3=\twelveex
      \scriptscriptfont3=\tenex
    \textfont\itfam=\fourteenit     \scriptfont\itfam=\tenit
    \textfont\slfam=\fourteensl     \scriptfont\slfam=\tensl
    \textfont\bffam=\fourteenbf     \scriptfont\bffam=\tenbf
      \scriptscriptfont\bffam=\sevenbf
    \textfont\ttfam=\fourteentt
    \textfont\cpfam=\fourteencp
    \samef@nt
    \b@gheight=14pt
    \setbox\strutbox=\hbox{\vrule height 0.85\b@gheight
				depth 0.35\b@gheight width\z@ }}
\def\twelvepoint{\relax
    \textfont0=\twelverm          \scriptfont0=\ninerm
      \scriptscriptfont0=\sixrm
    \textfont1=\twelvei           \scriptfont1=\ninei
      \scriptscriptfont1=\sixi
    \textfont2=\twelvesy           \scriptfont2=\ninesy
      \scriptscriptfont2=\sixsy
    \textfont3=\twelveex          \scriptfont3=\tenex
      \scriptscriptfont3=\tenex
    \textfont\itfam=\twelveit     \scriptfont\itfam=\nineit
    \textfont\slfam=\twelvesl     \scriptfont\slfam=\ninesl
    \textfont\bffam=\twelvebf     \scriptfont\bffam=\ninebf
      \scriptscriptfont\bffam=\sixbf
    \textfont\ttfam=\twelvett
    \textfont\cpfam=\twelvecp
    \samef@nt
    \b@gheight=12pt
    \setbox\strutbox=\hbox{\vrule height 0.85\b@gheight
				depth 0.35\b@gheight width\z@ }}
\def\tenpoint{\relax
    \textfont0=\tenrm          \scriptfont0=\sevenrm
      \scriptscriptfont0=\fiverm
    \textfont1=\teni           \scriptfont1=\seveni
      \scriptscriptfont1=\fivei
    \textfont2=\tensy          \scriptfont2=\sevensy
      \scriptscriptfont2=\fivesy
    \textfont3=\tenex          \scriptfont3=\tenex
      \scriptscriptfont3=\tenex
    \textfont\itfam=\tenit     \scriptfont\itfam=\seveni
    \textfont\slfam=\tensl     \scriptfont\slfam=\sevenrm
    \textfont\bffam=\tenbf     \scriptfont\bffam=\sevenbf
      \scriptscriptfont\bffam=\fivebf
    \textfont\ttfam=\tentt
    \textfont\cpfam=\tencp
    \samef@nt
    \b@gheight=10pt
    \setbox\strutbox=\hbox{\vrule height 0.85\b@gheight
				depth 0.35\b@gheight width\z@ }}
%
%
%
\normalbaselineskip = 20pt plus 0.2pt minus 0.1pt
\normallineskip = 1.5pt plus 0.1pt minus 0.1pt
\normallineskiplimit = 1.5pt
\newskip\normaldisplayskip
\normaldisplayskip = 20pt plus 5pt minus 10pt
\newskip\normaldispshortskip
\normaldispshortskip = 6pt plus 5pt
\newskip\normalparskip
\normalparskip = 6pt plus 2pt minus 1pt
\newskip\skipregister
\skipregister = 5pt plus 2pt minus 1.5pt
\newif\ifsingl@    \newif\ifdoubl@
\newif\iftwelv@    \twelv@true
\def\singlespace{\singl@true\doubl@false\spaces@t}
\def\doublespace{\singl@false\doubl@true\spaces@t}
\def\normalspace{\singl@false\doubl@false\spaces@t}
\def\Tenpoint{\tenpoint\twelv@false\spaces@t}
\def\Twelvepoint{\twelvepoint\twelv@true\spaces@t}
\def\spaces@t{\relax
      \iftwelv@ \ifsingl@\subspaces@t3:4;\else\subspaces@t1:1;\fi
       \else \ifsingl@\subspaces@t3:5;\else\subspaces@t4:5;\fi \fi
      \ifdoubl@ \multiply\baselineskip by 5
         \divide\baselineskip by 4 \fi }
\def\subspaces@t#1:#2;{
      \baselineskip = \normalbaselineskip
      \multiply\baselineskip by #1 \divide\baselineskip by #2
      \lineskip = \normallineskip
      \multiply\lineskip by #1 \divide\lineskip by #2
      \lineskiplimit = \normallineskiplimit
      \multiply\lineskiplimit by #1 \divide\lineskiplimit by #2
      \parskip = \normalparskip
      \multiply\parskip by #1 \divide\parskip by #2
      \abovedisplayskip = \normaldisplayskip
      \multiply\abovedisplayskip by #1 \divide\abovedisplayskip by #2
      \belowdisplayskip = \abovedisplayskip
      \abovedisplayshortskip = \normaldispshortskip
      \multiply\abovedisplayshortskip by #1
        \divide\abovedisplayshortskip by #2
      \belowdisplayshortskip = \abovedisplayshortskip
      \advance\belowdisplayshortskip by \belowdisplayskip
      \divide\belowdisplayshortskip by 2
      \smallskipamount = \skipregister
      \multiply\smallskipamount by #1 \divide\smallskipamount by #2
      \medskipamount = \smallskipamount \multiply\medskipamount by 2
      \bigskipamount = \smallskipamount \multiply\bigskipamount by 4 }
\def\normalbaselines{ \baselineskip=\normalbaselineskip
   \lineskip=\normallineskip \lineskiplimit=\normallineskip
   \iftwelv@\else \multiply\baselineskip by 4 \divide\baselineskip by 5
     \multiply\lineskiplimit by 4 \divide\lineskiplimit by 5
     \multiply\lineskip by 4 \divide\lineskip by 5 \fi }
\Twelvepoint  
\interlinepenalty=50
\interfootnotelinepenalty=5000
\predisplaypenalty=9000
\postdisplaypenalty=500
\hfuzz=1pt
\vfuzz=0.2pt
\voffset=0pt
\dimen\footins=8 truein
%
%
%
\def\pagecontents{
   \ifvoid\topins\else\unvbox\topins\vskip\skip\topins\fi
   \dimen@ = \dp255 \unvbox255
   \ifvoid\footins\else\vskip\skip\footins\footrule\unvbox\footins\fi
   \ifr@ggedbottom \kern-\dimen@ \vfil \fi }
\def\makeheadline{\vbox to 0pt{ \skip@=\topskip
      \advance\skip@ by -12pt \advance\skip@ by -2\normalbaselineskip
      \vskip\skip@ \line{\vbox to 12pt{}\the\headline} \vss
      }\nointerlineskip}
\def\makefootline{\baselineskip = 1.5\normalbaselineskip
                 \line{\the\footline}}
\newif\iffrontpage
\newif\ifletterstyle
\newif\ifp@genum
\def\nopagenumbers{\p@genumfalse}
\def\pagenumbers{\p@genumtrue}
\pagenumbers
\newtoks\paperheadline
\newtoks\letterheadline
\newtoks\paperfootline
\newtoks\letterfootline
\newtoks\letterinfo
\newtoks\Letterinfo
\newtoks\date
\footline={\ifletterstyle\the\letterfootline\else\the\paperfootline\fi}
\paperfootline={\hss\iffrontpage\else\ifp@genum\tenrm\folio\hss\fi\fi}
\letterfootline={\iffrontpage\LETTERFOOT\else\hfil\fi}
\Letterinfo={\hfil}
\letterinfo={\hfil}
\def\LETTERFOOT{\hfil} 
%
\def\LETTERHEAD{\vtop{\baselineskip=20pt\hbox to
\hsize{\hfil\seventeenrm\strut
CALIFORNIA INSTITUTE OF TECHNOLOGY \hfil}
\hbox to \hsize{\hfil\ninerm\strut
CHARLES C. LAURITSEN LABORATORY OF HIGH ENERGY PHYSICS \hfil}
\hbox to \hsize{\hfil\ninerm\strut
PASADENA, CALIFORNIA 91125 \hfil}}}
\headline={\ifletterstyle\the\letterheadline\else\the\paperheadline\fi}
\paperheadline={\hfil}
\letterheadline{\iffrontpage \LETTERHEAD\else
    \rm \ifp@genum \hfil \folio\hfil\fi\fi}
\def\monthname{\relax\ifcase\month 0/\or January\or February\or
   March\or April\or May\or June\or July\or August\or September\or
   October\or November\or December\else\number\month/\fi}
\def\today{\monthname\ \number\day, \number\year}
\date={\today}
\countdef\pageno=1      \countdef\pagen@=0
\countdef\pagenumber=1  \pagenumber=1
\def\advancepageno{\global\advance\pagen@ by 1
   \ifnum\pagenumber<0 \global\advance\pagenumber by -1
    \else\global\advance\pagenumber by 1 \fi \global\frontpagefalse }
\def\folio{\ifnum\pagenumber<0 \romannumeral-\pagenumber
           \else \number\pagenumber \fi }
\def\footrule{\dimen@=\prevdepth\nointerlineskip
   \vbox to 0pt{\vskip -0.25\baselineskip \hrule width 0.35\hsize \vss}
   \prevdepth=\dimen@ }
\newtoks\foottokens
\foottokens={}
\newdimen\footindent
\footindent=24pt
\def\vfootnote#1{\insert\footins\bgroup
   \interlinepenalty=\interfootnotelinepenalty \floatingpenalty=20000
   \singl@true\doubl@false\Tenpoint
   \splittopskip=\ht\strutbox \boxmaxdepth=\dp\strutbox
   \leftskip=\footindent \rightskip=\z@skip
   \parindent=0.5\footindent \parfillskip=0pt plus 1fil
   \spaceskip=\z@skip \xspaceskip=\z@skip
   \the\foottokens
   \Textindent{$ #1 $}\footstrut\futurelet\next\fo@t}
\def\Textindent#1{\noindent\llap{#1\enspace}\ignorespaces}
\def\footnote#1{\attach{#1}\vfootnote{#1}}

\let\footsymbol=\star
\newcount\lastf@@t           \lastf@@t=-1
\newcount\footsymbolcount    \footsymbolcount=0
\newif\ifPhysRev
\def\bumpfootsymbolcount{\relax
   \iffrontpage \bumpfootsymbolNP \else \advance\lastf@@t by 1
     \ifPhysRev \bumpfootsymbolPR \else \bumpfootsymbolNP \fi \fi
   \global\lastf@@t=\pagen@ }
\def\bumpfootsymbolNP{\ifnum\footsymbolcount <0 \global\footsymbolcount =0 \fi
    \ifnum\lastf@@t<\pagen@ \global\footsymbolcount=0
     \else \global\advance\footsymbolcount by 1 \fi }
\def\bumpfootsymbolPR{\ifnum\footsymbolcount >0 \global\footsymbolcount =0 \fi
      \global\advance\footsymbolcount by -1 }
\def\fd@f#1 {\xdef\footsymbol{\mathchar"#1 }}
\def\generatefootsymbol{\ifcase\footsymbolcount \fd@f 13F \or \fd@f 279
	\or \fd@f 27A \or \fd@f 278 \or \fd@f 27B \else
	\ifnum\footsymbolcount <0 \fd@f{023 \number-\footsymbolcount }
	 \else \fd@f 203 {\loop \ifnum\footsymbolcount >5
		\fd@f{203 \footsymbol } \advance\footsymbolcount by -1
		\repeat }\fi \fi }

\def\nonfrenchspacing{\sfcode`\.=3001 \sfcode`\!=3000 \sfcode`\?=3000
	\sfcode`\:=2000 \sfcode`\;=1500 \sfcode`\,=1251 }
\nonfrenchspacing
\newdimen\d@twidth
{\setbox0=\hbox{s.} \global\d@twidth=\wd0 \setbox0=\hbox{s}
	\global\advance\d@twidth by -\wd0 }
\def\removehglue{\loop \unskip \ifdim\lastskip >\z@ \repeat }
\def\roll@ver#1{\removehglue \nobreak \count255 =\spacefactor \dimen@=\z@
	\ifnum\count255 =3001 \dimen@=\d@twidth \fi
	\ifnum\count255 =1251 \dimen@=\d@twidth \fi
    \iftwelv@ \kern-\dimen@ \else \kern-0.83\dimen@ \fi
   #1\spacefactor=\count255 }
\def\step@ver#1{\relax \ifmmode #1\else \ifhmode
	\roll@ver{${}#1$}\else {\setbox0=\hbox{${}#1$}}\fi\fi }
\def\attach#1{\step@ver{\strut^{\mkern 2mu #1} }}
%
%
%
\newcount\chapternumber      \chapternumber=0
\newcount\sectionnumber      \sectionnumber=0
\newcount\equanumber         \equanumber=0
\let\chapterlabel=\relax
\let\sectionlabel=\relax
\newtoks\chapterstyle        \chapterstyle={\Number}
\newtoks\sectionstyle        \sectionstyle={\chapterlabel\Number}
\newskip\chapterskip         \chapterskip=\bigskipamount
\newskip\sectionskip         \sectionskip=\medskipamount
\newskip\headskip            \headskip=8pt plus 3pt minus 3pt
\newdimen\chapterminspace    \chapterminspace=15pc
\newdimen\sectionminspace    \sectionminspace=10pc
\newdimen\referenceminspace  \referenceminspace=25pc
\def\chapterreset{\global\advance\chapternumber by 1
   \ifnum\equanumber<0 \else\global\equanumber=0\fi
   \sectionnumber=0 \makechapterlabel}
\def\makechapterlabel{\let\sectionlabel=\relax
   \xdef\chapterlabel{\the\chapterstyle{\the\chapternumber}.}}
\def\alphabetic#1{\count255='140 \advance\count255 by #1\char\count255}
\def\Alphabetic#1{\count255='100 \advance\count255 by #1\char\count255}
\def\Roman#1{\uppercase\expandafter{\romannumeral #1}}
\def\roman#1{\romannumeral #1}
\def\Number#1{\number #1}
\def\BLANC#1{}
\def\titlestyle#1{\par\begingroup \interlinepenalty=9999
     \leftskip=0.02\hsize plus 0.23\hsize minus 0.02\hsize
     \rightskip=\leftskip \parfillskip=0pt
     \hyphenpenalty=9000 \exhyphenpenalty=9000
     \tolerance=9999 \pretolerance=9000
     \spaceskip=0.333em \xspaceskip=0.5em
     \iftwelv@\fourteenpoint\else\twelvepoint\fi
   \noindent #1\par\endgroup }
\def\spacecheck#1{\dimen@=\pagegoal\advance\dimen@ by -\pagetotal
   \ifdim\dimen@<#1 \ifdim\dimen@>0pt \vfil\break \fi\fi}
\def\TableOfContentEntry#1#2#3{\relax}
\def\chapter#1{\par \penalty-300 \vskip\chapterskip
   \spacecheck\chapterminspace
   \chapterreset \titlestyle{\chapterlabel\ #1}
   \TableOfContentEntry c\chapterlabel{#1}
   \nobreak\vskip\headskip \penalty 30000
   \wlog{\string\chapter\space \chapterlabel} }

\def\section#1{\par \ifnum\the\lastpenalty=30000\else
   \penalty-200\vskip\sectionskip \spacecheck\sectionminspace\fi
   \global\advance\sectionnumber by 1
   \xdef\sectionlabel{\the\sectionstyle\the\sectionnumber}
   \wlog{\string\section\space \sectionlabel}
   \TableOfContentEntry s\sectionlabel{#1}
   \noindent {\caps\enspace\sectionlabel\quad #1}\par
   \nobreak\vskip\headskip \penalty 30000 }
\def\subsection#1{\par
   \ifnum\the\lastpenalty=30000\else \penalty-100\smallskip \fi
   \noindent\undertext{#1}\enspace \vadjust{\penalty5000}}

\def\undertext#1{\vtop{\hbox{#1}\kern 1pt \hrule}}
\def\APPENDIX#1#2{\par\penalty-300\vskip\chapterskip
   \spacecheck\chapterminspace \chapterreset \xdef\chapterlabel{#1}
   \titlestyle{APPENDIX #2} \nobreak\vskip\headskip \penalty 30000
   \TableOfContentEntry a{#1}{#2}
   \wlog{\string\Appendix\ \chapterlabel} }
\def\Appendix#1{\APPENDIX{#1}{#1}}
\def\appendix{\APPENDIX{A}{}}
\def\unnumberedchapters{\let\makechapterlabel=\relax \let\chapterlabel=\relax
   \sectionstyle={\BLANC}\let\sectionlabel=\relax \sequentialequations }
%
%
%
\def\eqname#1{\relax \ifnum\equanumber<0
     \xdef#1{{\noexpand\rm(\number-\equanumber)}}%
       \global\advance\equanumber by -1
    \else \global\advance\equanumber by 1
      \xdef#1{{\noexpand\rm(\chapterlabel\number\equanumber)}} \fi #1}
\def\eqinsert#1{\noalign{\dimen@=\prevdepth \nointerlineskip
   \setbox0=\hbox to\displaywidth{\hfil #1}
   \vbox to 0pt{\kern 0.5\baselineskip\hbox{$\!\box0\!$}\vss}
   \prevdepth=\dimen@}}
%

%
%
\def\GENITEM#1;#2{\par \hangafter=0 \hangindent=#1
    \Textindent{$ #2 $}\ignorespaces}
\outer\def\newitem#1=#2;{\gdef#1{\GENITEM #2;}}
\newdimen\itemsize                \itemsize=30pt
\newitem\item=1\itemsize;
\newitem\sitem=1.75\itemsize;     
\newitem\ssitem=2.5\itemsize;     
\outer\def\newlist#1=#2&#3&#4;{\toks0={#2}\toks1={#3}%
   \count255=\escapechar \escapechar=-1
   \alloc@0\list\countdef\insc@unt\listcount     \listcount=0
   \edef#1{\par
      \countdef\listcount=\the\allocationnumber
      \advance\listcount by 1
      \hangafter=0 \hangindent=#4
      \Textindent{\the\toks0{\listcount}\the\toks1}}
   \expandafter\expandafter\expandafter
    \edef\c@t#1{begin}{\par
      \countdef\listcount=\the\allocationnumber \listcount=1
      \hangafter=0 \hangindent=#4
      \Textindent{\the\toks0{\listcount}\the\toks1}}
   \expandafter\expandafter\expandafter
    \edef\c@t#1{con}{\par \hangafter=0 \hangindent=#4 \noindent}
   \escapechar=\count255}
\def\c@t#1#2{\csname\string#1#2\endcsname}
\newlist\point=\Number&.&1.0\itemsize;
\newlist\subpoint=(\alphabetic&)&1.75\itemsize;
\newlist\subsubpoint=(\roman&)&2.5\itemsize;
%

%
%
%
%
\newcount\referencecount     \referencecount=0
\newcount\lastrefsbegincount \lastrefsbegincount=0
\newif\ifreferenceopen       \newwrite\referencewrite
\newif\ifrw@trailer
\newdimen\refindent     \refindent=30pt
\def\NPrefmark#1{\attach{\scriptscriptstyle [ #1 ] }}
\let\PRrefmark=\attach
\def\refmark#1{\relax\ifPhysRev\PRrefmark{#1}\else\NPrefmark{#1}\fi}
\def\refend@{\refmark{\number\referencecount}}
\def\refend{\refend@{}\space }
\def\refsend{\refmark{\count255=\referencecount
   \advance\count255 by-\lastrefsbegincount
   \ifcase\count255 \number\referencecount
   \or \number\lastrefsbegincount,\number\referencecount
   \else \number\lastrefsbegincount-\number\referencecount \fi}\space }
\def\refitem#1{\par \hangafter=0 \hangindent=\refindent \Textindent{#1}}
\def\Ref{\rw@trailertrue\REF}
\def\ref{\Ref\?}

\def\REF#1{\r@fstart{#1}%
   \rw@begin{\the\referencecount.}\rw@end}
\def\REFS#1{\r@fstart{#1}%
   \lastrefsbegincount=\referencecount
   \rw@begin{\the\referencecount.}\rw@end}
\def\r@fstart#1{\chardef\rw@write=\referencewrite \let\rw@ending=\refend@
   \ifreferenceopen \else \global\referenceopentrue
   \immediate\openout\referencewrite=referenc.txa
   \toks0={\catcode`\^^M=10}\immediate\write\rw@write{\the\toks0} \fi
   \global\advance\referencecount by 1 \xdef#1{\the\referencecount}}
{\catcode`\^^M=\active %
 \gdef\rw@begin#1{\immediate\write\rw@write{\noexpand\refitem{#1}}%
   \begingroup \catcode`\^^M=\active \let^^M=\relax}%
 \gdef\rw@end#1{\rw@@end #1^^M\rw@terminate \endgroup%
   \ifrw@trailer\rw@ending\global\rw@trailerfalse\fi }%
 \gdef\rw@@end#1^^M{\toks0={#1}\immediate\write\rw@write{\the\toks0}%
   \futurelet\n@xt\rw@test}%
 \gdef\rw@test{\ifx\n@xt\rw@terminate \let\n@xt=\relax%
       \else \let\n@xt=\rw@@end \fi \n@xt}%
}
\let\rw@ending=\relax
\let\rw@terminate=\relax
\let\splitout=\relax
\def\Closeout#1{\toks0={\catcode`\^^M=5}\immediate\write#1{\the\toks0}%
   \immediate\closeout#1}
%
%
\newcount\figurecount     \figurecount=0
\newcount\tablecount      \tablecount=0
\newif\iffigureopen       \newwrite\figurewrite
\newif\iftableopen        \newwrite\tablewrite
\def\FIG#1{\f@gstart{#1}%
   \rw@begin{\the\figurecount)}\rw@end}

\def\Fig{\rw@trailertrue\def\rw@ending{Fig.~\?}\FIG\?}
\def\fig{\rw@trailertrue\def\rw@ending{fig.~\?}\FIG\?}
\def\TABLE#1{\T@Bstart{#1}%
   \rw@begin{\the\tableecount:}\rw@end}
\def\Table{\rw@trailertrue\def\rw@ending{Table~\?}\TABLE\?}
\def\f@gstart#1{\chardef\rw@write=\figurewrite
   \iffigureopen \else \global\figureopentrue
   \immediate\openout\figurewrite=figures.txa
   \toks0={\catcode`\^^M=10}\immediate\write\rw@write{\the\toks0} \fi
   \global\advance\figurecount by 1 \xdef#1{\the\figurecount}}
\def\T@Bstart#1{\chardef\rw@write=\tablewrite
   \iftableopen \else \global\tableopentrue
   \immediate\openout\tablewrite=tables.txa
   \toks0={\catcode`\^^M=10}\immediate\write\rw@write{\the\toks0} \fi
   \global\advance\tablecount by 1 \xdef#1{\the\tablecount}}
%

%
%
%
\def\getfigure#1{\global\advance\figurecount by 1
   \xdef#1{\the\figurecount}\count255=\escapechar \escapechar=-1
   \edef\n@xt{\noexpand\g@tfigure\csname\string#1Body\endcsname}%
   \escapechar=\count255 \n@xt }
\def\g@tfigure#1#2 {\errhelp=\disabledfigures \let#1=\relax
   \errmessage{\string\getfigure\space disabled}}
\newhelp\disabledfigures{ Empty figure of zero size assumed.}
\def\figinsert#1{\midinsert\Tenpoint\medskip
   \count255=\escapechar \escapechar=-1
   \edef\n@xt{\csname\string#1Body\endcsname}
   \escapechar=\count255 \centerline{\n@xt}
   \bigskip\narrower\narrower
   \noindent{\it Figure}~#1.\quad }
%
%
%
\def\masterreset{\global\pagenumber=1 \global\chapternumber=0
   \global\equanumber=0 \global\sectionnumber=0
   \global\referencecount=0 \global\figurecount=0 \global\tablecount=0 }
\def\FRONTPAGE{\ifvoid255\else\vfill\penalty-20000\fi
      \masterreset\global\frontpagetrue
      \global\lastf@@t=0 \global\footsymbolcount=0}

\def\paperstyle{\letterstylefalse\normalspace\papersize}
\def\letterstyle{\letterstyletrue\singlespace\lettersize}
\def\papersize{\hsize=35 truepc\vsize=50 truepc\hoffset=-2.51688 truepc
               \skip\footins=\bigskipamount}
\def\lettersize{\hsize=5.5 truein\vsize=8.25 truein\hoffset=.4875 truein
	\voffset=.3125 truein
   \skip\footins=\smallskipamount \multiply\skip\footins by 3 }
\paperstyle   
%
%
\def\MEMO{\letterstyle \letterinfo={\hfil } \let\rule=\memorule
	\FRONTPAGE \memohead }
\let\memohead=\relax

\def\memit@m#1{\smallskip \hangafter=0 \hangindent=1in
      \Textindent{\caps #1}}
\def\subject{\memit@m{Subject:}}
\def\topic{\memit@m{Topic:}}
\def\from{\memit@m{From:}}
\def\to{\relax \ifmmode \rightarrow \else \memit@m{To:}\fi }
\def\memorule{\medskip\hrule height 1pt\bigskip}
\newwrite\labelswrite
\newtoks\rw@toks

\def\addressee#1{\null\vskip .5truein\line{
\hskip 0.5\hsize minus 0.5\hsize\the\date\hfil}\bigskip
   \ialign to\hsize{\strut ##\hfil\tabskip 0pt plus \hsize \cr #1\crcr}
   \writelabel{#1}\medskip\par\noindent}
\def\rwl@begin#1\cr{\rw@toks={#1\crcr}\relax
   \immediate\write\labelswrite{\the\rw@toks}\futurelet\n@xt\rwl@next}
\def\rwl@next{\ifx\n@xt\rwl@end \let\n@xt=\relax
      \else \let\n@xt=\rwl@begin \fi \n@xt}
\let\rwl@end=\relax
\def\writelabel#1{\immediate\write\labelswrite{\noexpand\labelbegin}
     \rwl@begin #1\cr\rwl@end
     \immediate\write\labelswrite{\noexpand\labelend}}
\newbox\FromLabelBox

\let\labelend=\relax
\def\labelbegin#1\labelend{\setbox0=\vbox{\ialign{##\hfil\cr #1\crcr}}
     \MakeALabel }
\newtoks\FromAddress
\FromAddress={}
\def\MakeFromBox#1{\global\setbox\FromLabelBox=\vbox{\Tenpoint
     \ialign{##\hfil\cr #1\the\FromAddress\crcr}}}
\newdimen\labelwidth		\labelwidth=6in
\def\MakeALabel{\vskip 1pt \hbox{\vrule \vbox{
	\hsize=\labelwidth \hrule\bigskip
	\leftline{\hskip 1\parindent \copy\FromLabelBox}\bigskip
	\centerline{\hfil \box0 } \bigskip \hrule
	}\vrule } \vskip 1pt plus 1fil }
\newskip\signatureskip       \signatureskip=30pt
\def\signed#1{\par \penalty 9000 \medskip \dt@pfalse
  \everycr={\noalign{\ifdt@p\vskip\signatureskip\global\dt@pfalse\fi}}
  \setbox0=\vbox{\singlespace \ialign{\strut ##\hfil\crcr
   \noalign{\global\dt@ptrue}#1\crcr}}
  \line{\hskip 0.5\hsize minus 0.5\hsize \box0\hfil} \medskip }
\newbox\letterb@x
\def\lettertext{\par\unvcopy\letterb@x\par}
\def\multiletter{\setbox\letterb@x=\vbox\bgroup
      \everypar{\vrule height 1\baselineskip depth 0pt width 0pt }
      \singlespace \topskip=\baselineskip }
\def\letterend{\par\egroup}
%
%
%
\newskip\frontpageskip
\newtoks\Pubnum
\newtoks\pubtype
\newif\ifp@bblock  \p@bblocktrue
\def\PH@SR@V{\doubl@true \baselineskip=24.1pt plus 0.2pt minus 0.1pt
             \parskip= 3pt plus 2pt minus 1pt }
\def\PHYSREV{\paperstyle\PhysRevtrue\PH@SR@V}
\def\titlepage{\FRONTPAGE\paperstyle\ifPhysRev\PH@SR@V\fi
   \ifp@bblock\p@bblock \else\hrule height\z@ \relax \fi }
\def\nopubblock{\p@bblockfalse}

\frontpageskip=12pt plus .5fil minus 2pt
\pubtype={\tensl Preliminary Version}
\Pubnum={}
\def\p@bblock{\begingroup \tabskip=\hsize minus \hsize
   \baselineskip=1.5\ht\strutbox \topspace-2\baselineskip
   \halign to\hsize{\strut ##\hfil\tabskip=0pt\crcr
       \the\Pubnum\crcr\the\date\crcr\the\pubtype\crcr}\endgroup}
\def\title#1{\vskip\frontpageskip \titlestyle{#1} \vskip\headskip }
\def\author#1{\vskip\frontpageskip\titlestyle{\twelvecp #1}\nobreak}

\def\address#1{\par\kern 5pt\titlestyle{\twelvepoint\it #1}}
\def\andaddress{\par\kern 5pt \centerline{\sl and} \address}

\def\abstract{\par\dimen@=\prevdepth \hrule height\z@ \prevdepth=\dimen@
   \vskip\frontpageskip\centerline{\fourteenrm ABSTRACT}\vskip\headskip }

%
%
%

\def\\{\relax \ifmmode \backslash \else {\tt\char`\\}\fi }
\def\sequentialequations{\relax\if\equanumber<0\else\global\equanumber=-1\fi}

\def\journal#1&#2(#3){\unskip, \sl #1\unskip~\bf\ignorespaces #2\rm (19#3),}

\def\topspace{\hrule height 0pt depth 0pt \vskip}

\def\Buildrel#1\under#2{\mathrel{\mathop{#2}\limits_{#1}}}
\def\becomes#1{\mathchoice{\becomes@\scriptstyle{#1}}{\becomes@\scriptstyle
   {#1}}{\becomes@\scriptscriptstyle{#1}}{\becomes@\scriptscriptstyle{#1}}}
\def\becomes@#1#2{\mathrel{\setbox0=\hbox{$\m@th #1{\,#2\,}$}%
	\mathop{\hbox to \wd0 {\rightarrowfill}}\limits_{#2}}}

\let\int=\intop         
\def\lsim{\mathrel{\mathpalette\@versim<}}
\def\gsim{\mathrel{\mathpalette\@versim>}}
\def\@versim#1#2{\vcenter{\offinterlineskip
	\ialign{$\m@th#1\hfil##\hfil$\crcr#2\crcr\sim\crcr } }}
\def\big#1{{\hbox{$\left#1\vbox to 0.85\b@gheight{}\right.\n@space$}}}
\def\Big#1{{\hbox{$\left#1\vbox to 1.15\b@gheight{}\right.\n@space$}}}
\def\bigg#1{{\hbox{$\left#1\vbox to 1.45\b@gheight{}\right.\n@space$}}}
\def\Bigg#1{{\hbox{$\left#1\vbox to 1.75\b@gheight{}\right.\n@space$}}}
%
%
%
\let\sec@nt=\sec
\def\sec{\relax\ifmmode\let\n@xt=\sec@nt\else\let\n@xt\section\fi\n@xt}
\def\obsolete#1{\message{Macro \string #1 is obsolete.}}
\def\firstsec#1{\obsolete\firstsec \section{#1}}
\def\firstsubsec#1{\obsolete\firstsubsec \subsection{#1}}
\def\thispage#1{\obsolete\thispage \global\pagenumber=#1\frontpagefalse}
\def\thischapter#1{\obsolete\thischapter \global\chapternumber=#1}
\def\REFSCON{\obsolete\REFSCON\REF}
\def\splitout{\obsolete\splitout\relax}
\def\prop{\obsolete\prop \propto }
\def\nextequation#1{\obsolete\nextequation \global\equanumber=#1
   \ifnum\the\equanumber>0 \global\advance\equanumber by 1 \fi}
\def\BOXITEM{\afterassigment\B@XITEM\setbox0=}
\def\B@XITEM{\par\hangindent\wd0 \noindent\box0 }
\def\phyzzx{PHY\setbox0=\hbox{Z}\copy0 \kern-0.5\wd0 \box0 X}
%
%
\everyjob{\xdef\today{\monthname\ \number\day, \number\year}}
        
%


\hoffset=0.2truein
\voffset=0.1truein
\hsize=6truein
\def\TITLEPAGE{\frontpagetrue}
\def\CALT#1{\hbox to\hsize{\tenpoint \baselineskip=12pt
	\hfil\vtop{\hbox{\strut CALT-68-#1}
	\hbox{\strut DOE RESEARCH AND}
	\hbox{\strut DEVELOPMENT REPORT}}}}

\def\CALTECH{\smallskip
	\address{California Institute of Technology, Pasadena, CA 91125}}
\def\TITLE#1{\vskip 1in \centerline{\fourteenpoint #1}}
\def\AUTHOR#1{\vskip .5in \centerline{#1}}

\def\ABSTRACT#1{\vskip .5in \vfil \centerline{\twelvepoint \bf Abstract}
	#1 \vfil}
\def\ENDTITLEPAGE{\vfil\eject\pageno=1}

\def\sqr#1#2{{\vcenter{\hrule height.#2pt
      \hbox{\vrule width.#2pt height#1pt \kern#1pt
        \vrule width.#2pt}
      \hrule height.#2pt}}}

\def\section#1#2{
\noindent\hbox{\hbox{\bf #1}\hskip 10pt\vtop{\hsize=5in
\baselineskip=12pt \noindent \bf #2 \hfil}\hfil}
\medskip}

\def\underwig#1{	
	\setbox0=\hbox{\rm \strut}
	\hbox to 0pt{$#1$\hss} \lower \ht0 \hbox{\rm \char'176}}

\def\bunderwig#1{	
	\setbox0=\hbox{\rm \strut}
	\hbox to 1.5pt{$#1$\hss} \lower 12.8pt
	 \hbox{\seventeenrm \char'176}\hbox to 2pt{\hfil}}

\def\MEMO#1#2#3#4#5{
\frontpagetrue
\centerline{\tencp INTEROFFICE MEMORANDUM}
\smallskip
\centerline{\bf CALIFORNIA INSTITUTE OF TECHNOLOGY}
\bigskip
\vtop{\tenpoint \hbox to\hsize{\strut \hbox to .75in{\caps to:\hfil}
\hbox to3in{#1\hfil}
\hbox to .75in{\caps date:\hfil}\quad \the\date\hfil}
\hbox to\hsize{\strut \hbox to.75in{\caps from:\hfil}\hbox to 2in{#2\hfil}
\hbox{{\caps extension:}\quad#3\qquad{\caps mail code:\quad}#4}\hfil}
\hbox{\hbox to.75in{\caps subject:\hfil}\vtop{\parindent=0pt
\hsize=3.5in #5\hfil}}
\hbox{\strut\hfil}}}

\def\box{{\hbox{ $\sqcup$}\llap{\hbox{$\sqcap$}}}}
\def\hatbox{{\hbox{ $\sqcup$}\llap{\hbox{$\hat\sqcap$}}}}
\TITLEPAGE
\CALT{1745}
\TITLE{Path Integral Over Conformally Self--Dual Geometries
\footnote\dagger{Work supported in part by the U.S. Department of Energy
under Contract No. DE-AC 0381-ER40050.\vskip 0mm}}
\AUTHOR{Christof Schmidhuber\footnote*{christof@theory3.caltech.edu}}
\CALTECH
\ABSTRACT{
The path integral of four dimensional quantum gravity
is restricted to conformally
self-dual metrics.
It reduces to integrals
over the conformal
factor and over the moduli space of conformally self--dual metrics
and can be studied with the methods of two dimensional quantum
gravity in conformal gauge.
The conformal anomaly induces an analog of the
Liouville action.
The proposal of David, Distler and Kawai is generalized to
four dimensions. Critical exponents and the analog of the $c=1$ barrier
of two dimensional gravity are derived.
Connections with Weyl gravity and four dimensional
topological gravity are suggested.
\vfill
hepth@xxx/9112005
}
\ENDTITLEPAGE

{\leftline{\bf 1. Introduction}}

The path integral of two dimensional quantum gravity
in conformal gauge essentially
reduces to an integral over the conformal factor. Its dynamics are provided
by the conformal anomaly which can be integrated
to give the Liouville action.
\Ref\poy{A.M. Polyakov, Phys. Lett. 103B, 207 (1981)}
David, Distler and Kawai (DDK) proposed reformulating the theory
in terms of free fields propagating
in a fictitious gravitational background with a marginal operator
in the role of the cosmological constant term.
\Ref\d{F. David, Mod. Phys. Lett. A3, 1651 (1988)}
 \Ref\dk{J. Distler and H. Kawai, Nucl. Phys. B 321, 509 (1988)}
The scaling laws derived from this proposal, e.g. for the fixed area
partition function $Z(A)$ at zero cosmological constant,
are in agreement with the results from light cone gauge,
\Ref\kpz{V. Knizhnik, A. M. Polyakov and A. B. Zamolodchikov,
Mod. Phys. Lett. A3, 819 (1988)}
computer simulations\Ref\mag{M. E. Agishtein and A. A. Migdal,
Nucl. Phys. B350, 690 (1991)}
and matrix models.\Ref\mxm{E.g., D. J. Gross and A. A. Migdal,
Nucl. Phys. B 340, 333 (1990) and references therein.}
These laws have
been derived for two dimensional quantum gravity coupled to conformal
field theories with $c\le 1$, including nonunitary ones.

It is natural to ask whether we can apply the same methods to four
dimensional quantum gravity and, e.g., derive the analog of the $c=1$
barrier or scaling laws
that could be compared with computer simulations. In this paper I show
that this is the case at least if we restrict ourselves to conformally
self-dual metrics:
the path integral to study is
$$\int Dg\ Dx\ Dp\ e^{- S_{mat}[g,x] - \int_{M} d^4 x{\sqrt g}(\lambda +
 \gamma R +\eta R^2 +ip W_+)}\ O_1[g,x]...O_n[g,x]\eqno(1.1)$$
with a Lagrange multiplier $p$.
The manifold M has fixed topology and Euclidean signature.
$x$ represents some conformally invariant matter
fields, $O_1...O_n$ are local operators, $\lambda$ and $\gamma$ are
the cosmological and inverse Newtonian
constants. The $R^2$-term will be needed to cancel an $R^2$-term from the
conformal anomaly~and~will disappear in the end.
$W_{\pm}$ is the (anti-) self-dual part of the (traceless) Weyl tensor
$$\eqalign{&W_{\pm\ \mu\nu\sigma\tau}\equiv {1\over 2}
( W_{\mu\nu\sigma\tau} \pm
 {1\over 2}\epsilon_{\mu\nu}^{\ \ \ \alpha\beta} W_{\alpha\beta\sigma\tau}),\cr
&W_{\mu\nu\sigma\tau}=R_{\mu\nu\sigma\tau}
-{1\over 2}(g_{\mu\sigma}R_{\nu\tau}+g_{\nu\tau}R_{\mu\sigma}
-g_{\mu\tau}R_{\nu\sigma}-g_{\nu\sigma}R_{\mu\tau})
+{1\over6}(g_{\mu\sigma}g_{\nu\tau}-g_{\mu\tau}g_{\nu\sigma})R.}$$
The Lagrange multiplier p is a $4^{th}$ rank self-dual tensor
field which (like $W_+$) transforms as (2,0)
under the Euclideanized Lorentz group $SO(4)\sim SU(2)\times SU(2)$,
i.e., like a spin 2 field.

The merit of $p$ is that it reduces
the path integral over
the metric to an integral over the
finite dimensional moduli space of conformally (anti-) self-dual metrics,
the conformal factor and the diffeomorphism group, as will
be explained. This
makes (1.1) similar to two dimensional quantum gravity,
the moduli space of conformally self-dual
metrics playing the role of the moduli space of Riemann surfaces.
As a consequence, we will be able to apply the same methods.

Why is ``conformally (anti-) self-dual quantum gravity"
interesting? One reason is that it is a four dimensional model
for gravity that we might be able to quantize consistently and solve
and in which we can nicely study, e.g., the effects of the conformal anomaly.
It is a much richer toy model than two dimensional
quantum gravity, because there the one dimensional universes have no geometry,
only a total
length, and the wave function of the universe is therefore relatively trivial.
Conformally self-dual gravity will also be argued
to describe renormalization
group fixed points of gravity with a Weyl term,
characterized by infinite Weyl coupling
constant.
One might speculate that
they correspond to long or short distance ``phases".
One might also hope that some of the results found in this infinite
coupling limit prevail qualitatively at finite or zero Weyl coupling.

Of course, there is a well known ghost problem common to all fourth
order derivative actions (like the Weyl action:)
we can rewrite them in terms of new fields with two
derivatives only, but some of them will have the wrong
sign in the kinetic term. With Minkowskian signature this leads
to nonunitarity.
Two types of fourth order derivative terms will arise in this paper:
terms quadratic in the curvature, and a new term induced by the
conformal anomaly which makes the model renormalizable.
The former will either be tuned away in the end
or decouple. The ghosts arising from the latter might decouple due to
the reparametrization constraints, as in string theory. In any case,
they will not affect the calculations done.

Part of this paper is concerned with the four
dimensional analog of the Liouville action and of DDK.
In a different context the induced action for the conformal factor
and its renormalization have also been studied recently
by Antoniadis and Mottola. I have used some of their calculations.
\Ref\ma{I. Antoniadis and E. Mottola, Los Alamos preprint LA-UR-91-1653 (1991)}
However, when I discuss the four dimensional analog of DDK's
method of decoupling the conformal factor from its measure, my treatment
and my conclusions will differ from those of [\ma]. I will state the main
differences.

In section two, (1.1) without the operators will be
rewritten as an integral over moduli space and over the conformal factor
$\phi$ with a few determinants
in its gravitational background. As in two dimensions,
the determinants can be decoupled from $\phi$
by introducing a 4D analog of the Liouville action.
Its form has already been found in [\ma]. It consists of a free $4^{th}$ order
derivative piece (essentially $\phi\box^2\phi$) plus pieces that renormalize
$\lambda, \gamma$ and $\eta$ in (1.1), as explained in section 3.

In quantum gravity
the measure $D_g g$ for the metric is defined with respect to
the fluctuating metric itself.
In two dimensions, DDK
(further elaborated in \Ref\dho{E. D' Hoker, Mod. Phys. Lett. A6, 745 (1991)})
replaced the conformal factor by a field whose measure is defined with respect
to some background metric and whose action is again the Liouville
action with modified coefficients, determined by requiring
that the choice for the background metric is irrelevant.
In chapter 4, the same is done in four dimensions.
The cosmological constant, the Hilbert-Einstein term and the $R^2$
term should each become truly marginal operators of the new theory,
but so far I have verified this only for the cosmological term.

In section 5 scaling laws in
conformally self-dual quantum
gravity are derived, similar to the two dimensional ones.
\footnote*{the values of the exponents and the analog of the
$c=1$ barrier can be found in the note added
at the end of the paper}
I will discuss the partition function at fixed volume or average curvature,
and the correlation
functions of local operators in their dependence on the cosmological constant.
It would be very interesting
to explore whether the condition $W_+=0$ can be imposed in
computer simulations of random triangulations. Then these predictions
could be compared with ``experiment."

Section 6 discusses the connection of (1.1) with
fixed points of gravity with a Weyl term.
It is also suggested that conformally self-dual gravity is
connected with four dimensional topological gravity,
\Ref\wiw{E. Witten, Phys. Lett. B 206, 601 (1988);
J.M.F. Labastida and M. Pernici, Phys. Lett. B 212, 56 (1988)}
as in the two dimensional case.
\Ref\wv{E. Witten, Nucl. Phys. B 340, 281 (1990); E. and H. Verlinde,
Nucl. Phys. B 348, 457 (1991)}
\vfill
\eject

{\bf{2. Conformal Gauge}}

The Lagrange multiplier p in (1.1) restricts
the path integral over g
to conformally self-dual metrics, i.e. metrics with $W_+=0$.
$W_+$ has five independent components and the condition $W_+=0$ is Weyl-
and diffeomorphism invariant. So up to a finite number of moduli
the five surviving components of the metric will be the
conformal factor and the diffeomorphisms.
Let $m_i$ parametrize the moduli space of conformally self-dual
metrics modulo diffeomorphisms $x \rightarrow x+\xi$ and Weyl
transformations $g \rightarrow g e^\phi$. Let us
fix a representative $\hat g (m_i)$ via, say, the condition
$\hat R=0$ and Lorentz gauge
$\partial^\mu \hat g_{\mu\nu}=0$
and let us pick a conformally self dual metric
$$g_0 = (\hat g(m_i) e^\phi )^\xi\eqno(2.1)$$ where $\xi$ indicates
the action of a diffeomorphism.
At $g_0$ we can split up g: $$\delta g_{\mu\nu} = g_{0\mu\nu}\delta \phi
+ \nabla_{(\mu}\delta\xi_{\nu)}+\delta \bar h_{\mu\nu}.$$
The four $\delta\xi$'s generate infinitesimal diffeomorphisms and the
five $\bar h_{\mu\nu}$ parametrize the space of metrics
perpendicular to $\xi$,  $\phi$ and the moduli, i.e., perpendicular
to the conformally self dual ones.
The measure for g is defined, in analogy to two dimensions${}^{[\poy]}$, by
$$\Vert \delta g \Vert ^2 \equiv \int d^4 x {\sqrt g}(4
(\delta \phi+{1\over 2}\nabla^\mu \delta\xi_\mu)^2
 + (L\delta\xi)^2 + (\delta \bar h)^2)\eqno (2.2)$$ with
$$(L\delta\xi)_{\mu\nu} \equiv \nabla_{(\mu} \delta\xi_{\nu )} -
 {1 \over 2}g_{\mu\nu} \nabla^\rho \delta\xi_\rho.\eqno(2.3)$$
Apart from restricting the path integral, integrating out $p$ and $\bar h$
in (1.1) will contribute the determinant
$$\det(O^\dagger O)_
{g_0}^{-{1\over 2}}\eqno(2.4) $$
where $O^\dagger$ is the linearized $W_+$-term
$$(O^\dagger_{g_0} \bar h)_{\mu\nu\sigma\tau} \equiv\lim_{\epsilon
\rightarrow 0} {1\over \epsilon}( W_{+\ \mu\nu\sigma\tau}
[g_0 +\epsilon \bar h] -W_{+\ \mu\nu\sigma\tau}[g_0]),\eqno(2.5)$$
$O$ is its adjoint and $O^\dagger O$ is a $4^{th}$
order, conformally invariant,
linear differential operator in the curved background $g_0$, acting on $p$.

We are left with an integral over the conformal
equivalence class of each $\hat g$.
{}From (2.2) it is seen that changing variables from $g$ to $\phi$
and $\xi$ in this equivalence class leads to a Jacobian
$$\det (L^\dagger L)_g^{1\over 2}$$
where the zero modes of the operator $L$, defined in
(2.3), have to be projected out.
After dropping the integral over the diffeomorphism group $D\xi$ (since
gravitational anomalies can occur only in 4k+2 dimensions
\Ref\awi{L. Alvarez-Gaum\'e and E. Witten, Nucl. Phys. B 234, 269 (1983)})
the path integral (1.1) without the operator insertions reduces
to an integral over the
moduli space of conformally self-dual metrics and $\phi$:
$$ \int \prod_i dm_i\ D\phi\
\det (O^\dagger O)^{-{1\over 2}}_{\hat g e^\phi}\
\det( L^\dagger L)^{1\over 2}_{\hat g e^\phi}\
\det( \triangle)^{-{1\over 2}}_{\hat g e^\phi}\
e^{-\int d^4 x {\sqrt g}(\lambda+\gamma R+\eta R^2)} \eqno (2.6)$$
where the matter partition function has been
denoted by $\det(\triangle)^{-{1\over 2}}$. Despite of the notation, the
conformally invariant matter is allowed to be fermions,
Yang-Mills fields, etc., as well as conformally coupled scalars.

The moduli space of conformally self-dual metrics
is a very interesting subject by itself
which will not be discussed here. On the four sphere
its dimension is zero: all conformally self-dual metrics on $S^4$
are conformally flat.
On $K^3$, e.g., its dimension is 57.
\Ref\egu{In T. Eguchi, P. B. Gilkey and
A. J. Hanson, Phys. Rep. 66, 213 (1988), after I. M. Singer}

\vfill
\eject

{\bf{3. Liouville in 4D}}

Let us now decouple the determinants in (2.6)
from $\phi$. For conformally invariant differential operators $X$:
\footnote*{More precisely, if $X\equiv M^\dagger M$,
$M$ has to transform as $e^{p\phi}Me^{q\phi}$ under
$g\rightarrow ge^\phi$. It can be shown
that $L$ transforms as $e^{\phi}Le^{-\phi}$.}
$$\det X_{\hat g e^\phi} = \det X_{\hat g} e^{-S_i[\hat g,\phi]}\eqno(3.1)$$
where the induced action $S_i$ is obtained
from integrating the trace anomaly of the stress tensor
\Ref\bd{E.g., M. J. Duff, Nucl. Phys. B125, 334 (1977);
N. D. Birrell and P. C. W. Davies,
``Quantum Fields in Curved Space," Cambridge University Press 1982}
$$-2{{\delta S_i[\hat g,\phi]} \over{ \delta \phi}} ={\sqrt g}
<T^\mu _\mu> = {1\over{16\pi^2}}{\sqrt g}[a(F+{2 \over 3} \box R) + bG
]\ \ \ - 4\lambda'{\sqrt g} - 2\gamma'{\sqrt g} R \eqno (3.2)$$
where $F=W_+^2+W_-^2$ is the square of the Weyl tensor. (3.2) has,
apart from the divergent parameters $\lambda'$ and $ \gamma'$,
two finite parameters $a,b$.
${\sqrt g}G$ is the Gauss-Bonnet
density whose integral over the manifold
is proportional to the Euler characteristic.
Following Antoniadis and Mottola ${}^{[\ma]}$, (3.2) can actually
easily be integrated by noting that with $g=\hat ge^\phi$ the combination
$${\sqrt g}(G-{2\over 3}\box R)={\sqrt {\hat g}}\hat M
\ \phi + {\sqrt{\hat g}}(\hat G-{2\over 3}\hatbox\hat R)$$
is only linear in $\phi$ with the fourth order differential operator
$$\eqalign{\hat M &\equiv 2\hatbox^2 + 4\hat R^{\mu\nu}\hat\nabla_\mu\hat
\nabla_\nu-{4\over 3}\hat R\hatbox+{2\over 3}
(\hat\nabla^\mu\hat R)\hat\nabla_\mu\cr
&= 2\hatbox^2 + 4\hat R^{\mu\nu}
\hat\nabla_\mu\hat\nabla_\nu\ \ \hbox{if}\ \ \hat R=0\cr
&= 2\hatbox^2 \ \ \hbox{if}\ \ \hat g=\delta e^{\phi_0}.}\eqno(3.3)$$
${\sqrt g} F$ is independent of $\phi$ and ${\sqrt g}\box R$
integrates to the $R^2$ action. So the four dimensional analog
of the Liouville action consists of a free part plus
a cosmological constant term, a Hilbert-Einstein term and an $R^2$ term:

$$S_i[\hat g,\phi] =
{-b\over{32\pi^2}}S_0[\hat g,\phi]+{-a\over{32\pi^2}}S_1[\hat g,\phi]
 +{{a+b}\over{72\pi^2}}S_{R^2}+\gamma'S_R+\lambda'S_{c.c.}\eqno(3.4)$$ where
$$\eqalign{
S_0[\hat g,\phi] &= \int d^4 x {\sqrt{ \hat g}}
[ {1 \over 2} \phi \hat M \phi+(\hat G -{2\over 3}\hatbox\hat R)\phi]\cr
S_1[\hat g,\phi] &= \int d^4 x {\sqrt{ \hat g}} \hat F \phi\cr
S_{c.c} &= \int d^4x \sqrt{ \hat g} e^{2\phi}\cr
S_{R}&= \int d^4 x \sqrt{ \hat g} e^{\phi} [\hat R -{3\over 2}
(\hat \nabla \phi)^2 -3 \hatbox \phi]\cr
S_{R^2}&= \int d^4 x \sqrt{ \hat g}
[\hat R -{3\over 2}(\hat \nabla \phi)^2 -3 \hatbox \phi]^2}\eqno(3.5)$$
$b$ will turn out to be negative for `normal' operators $X$.
$\gamma', \lambda' $ and ${{a+b}\over{72\pi^2}}$
just renormalize $\gamma, \lambda $ and $\eta$.
A $\phi$-independent local term
$${-\int d^4 x\sqrt {\hat g}}({{a+b}\over{72\pi^2}}\hat R^2 +\gamma'
 \hat R+\lambda')\eqno(3.6)$$
has been omitted in (3.4) and will frequently be omitted in the following.
If it is included we see from (3.1) that for some action $S_j$:
$$S_i[\hat g,\phi]=S_j[\hat g e^\phi]-S_j[\hat g].\eqno(3.7)$$
${-b\over{32\pi^2}}S_0$ is the 4D analog of the 2D action
$$S_{2D}={c\over {48\pi}}\int d^4 x
{\sqrt {\hat g}}({1\over 2}\phi\hatbox\phi-\hat R\phi).$$
If $\hat g = \tilde g e^{\phi_0}$, $S_0$ can be written:
$$S_0[\hat g,\phi]=\int d^4x {\sqrt {\tilde g}}{1\over 2}
[(\phi+\phi_0)\tilde M(\phi+\phi_0)-\phi_0\tilde M\phi_0].\eqno(3.8)$$
Adding up the anomaly coefficients in (3.2) for
$\det(O^\dagger O)^{-{1\over2}},
\det(L^\dagger L)^{+{1\over2}},\det\Delta^{-{1\over2}}$,
$$A_0\equiv a_O +a_L+a_{mat}\hskip 1in
 B_0\equiv b_O+b_L+b_{mat},\eqno(3.9)$$
(2.6) can now be rewritten as

$$\eqalign{\int \prod_i dm_i\ &\chi(m_i)
\int D\phi\ e^{{B_0\over{32\pi^2}}S_0[\hat g ,\phi]+{A_0\over{32\pi^2}}
[\hat g, \phi]-\eta_1 S_{R^2}-\gamma_1S_R-\lambda_1S_{c.c.}}\cr
&\chi(m_i)\equiv
\det (O^\dagger O)^{-{1\over 2}}_{\hat g(m_i) }\
\det( L^\dagger L)^{1\over 2}_{\hat g(m_i)}\
\det (\triangle)^{-{1\over 2}}_{\hat g(m_i) }} \eqno(3.10)$$
$\chi(m_i)$ is now purely a function of the moduli $m_i$, once
we have fixed a representative $\hat g(m_i)$ for each point in moduli space.

The coefficients $a$ and $b$ in (3.2) have been calculated for, e.g.,
${}^{[\bd]}$

\line{conformally coupled
scalars ($\triangle \sim\box -{1\over 6}R$):\hfill
$a_0={1\over {120}}\hskip 1cm b_0=-{1\over{360}}$\hskip 2cm}
\line{spin ${1\over 2}$ (four component) fermions:\hfill
$a_{1\over 2}={6\over {120}}\hskip 1cm b_{1\over 2}=-{11\over{360}}$
\hskip .8cm(3.11)}
\line{massless gauge fields:\hfill
$a_1={12\over {120}}\hskip 1cm b_1=-{62\over{360}}$\hskip 2cm}

$a_O, a_L, b_O, b_L$, as well as
$a_M,b_M$ of $M$ in (3.3), which will also be needed later,
can in principle be calculated as usual
with the Schwinger-de Witt method.
\footnote*{see the note added at the end of the paper for the values.}
This will
be tedious, especially
for the fourth order derivative
operator $O^\dagger O$ which acts on a $4^{th}$ rank tensor field
and is the sum of many components.
It may be easier to study the operator product expansion of the
$flat$ $space$ stress tensor with itself and to see whether one can infer
the anomaly coefficients $a,b$ from it, as in
two dimensions. There the central charge c is read off from
$$T(z)T(w)\sim{{c\over 2}\over{(z-w)^2}}+...\ .$$

Note that the fourth order derivative induced action
makes the theory power counting renormalizable, and also bounded if
$b<0$.
The price is the existence of a ghost,
the general problem of fourth order derivative actions mentioned in
the introduction.
It has already been suggested in [\ma] that the reparametrization
constraints $T_{\mu\nu}\sim 0$ eliminate these ghosts from the physical
spectrum as they do in two dimensions.
\Ref\gsw{E.g., M.B. Green, J.H. Schwarz and E. Witten, ``Superstring
Theory", Cambridge University Press 1987}
This will be very interesting to explore in the future.
\vfill\eject

{\bf{4. David, Distler and Kawai in 4D}}

Let us now focus
on the $\phi$ integral over the conformal equivalence class of $\hat g$:
$$Z[\hat g]\equiv\int D_{\hat ge^\phi}\ \phi
\ e^{{B_0\over{32\pi^2}}S_0[\hat g ,\phi]+{A_0\over{32\pi^2}}
S_1[\hat g, \phi]-\eta_1 S_{R^2}-\gamma_1S_R-\lambda_1S_{c.c.}}\eqno(4.1)$$
where the dependence of
$Z$ on $\eta_1,\gamma_1$ and $\lambda_1$ has been suppressed.
In $D_{\hat ge^\phi}\phi$ it is indicated that the measure
for $\phi$ depends on $\phi$ itself, namely in two ways:
First, the metric itself must be used to define a norm in the space of metrics:
$$\Vert\delta\phi\Vert^2\equiv\int d^4 x{\sqrt g}
\ (\delta\phi(x))^2=\int d^4x{\sqrt{\hat g}}e^{2\phi(x)}(\delta\phi(x))^2.$$
Second, in order to define a short distance cutoff one should also
use the metric $\hat ge^\phi$ itself: the cutoff fluctuates with the field.
Let us follow David, Distler and Kawai ${}^{[\d],[\dk]}$ and assume
that the $\phi$-dependence of the measure in (4.1) can be absorbed
in a local renormalizable action:
$$D_{\hat g e^\phi}\ \phi\ e^{-S_i[\hat g,\phi]} =
D_{\hat g }\ \phi\ e^{-S_{{loc}}[\hat g,\phi]},\eqno(4.2)$$
where now on the right-hand side
$$\Vert\delta\phi\Vert^2\equiv\int d^4 x{\sqrt{\hat g}}(\delta\phi(x))^2$$
and the cutoff no longer fluctuates.

What is $S_{{loc}}$? Although the $\phi$- dependence of the measure
in (4.1) looks inconvenient we do learn something important
from (4.1): simultaneously changing
$$\hat g_{\mu\nu}\rightarrow\hat g_{\mu\nu}e^{\phi_0},
\hskip 1in \phi\rightarrow\phi-\phi_0$$
does not change the measure or $S_{R^2},S_R,S_{c.c.}$.
It does change the induced action. From (3.7) we see (reinstating the $\phi$-
independent terms (3.6) into (4.1)):
$$S_i[\hat g,\phi]\rightarrow S_i[\hat g,\phi]-S_i[\hat g,\phi_0]$$
We conclude
$$Z[\hat ge^{\phi_0}]=Z[\hat g]\ e^{S_i[\hat g,\phi_0]}\eqno(4.3)$$
that is, the $\phi$-theory behaves as if it
were a conformal field theory with conformal anomaly (3.2) given by
$a=-A_0,\ b=-B_0$. This is, of course, precisely what is needed
in order to insure that the background metric is really a fake:
if we vary it,
$\hat g_{\mu\nu}\rightarrow\hat g_{\mu\nu}e^{\phi_0}$, the variation
of the determinants in (3.10) is determined by their total conformal
anomalies $+A_0,\ +B_0$, defined in (3.9), and that just
cancels the $-A_0,\ -B_0$ from the $\phi$ theory.

So let us replace (4.1) as in (4.2) by a four dimensional
conformal field theory with conformal anomaly given by
$$a=-a_O-a_L-a_{{mat}},\hskip 1in b=-b_O-b_L-b_{{mat}}.\eqno(4.4)$$
I will propose -- and justify in a moment --
that as in two dimensions $S_{loc}$ in (4.2) is again the induced action
with modified coefficients $A,B$ and modified interactions:
$$Z[\hat g]\sim\int D_{\hat g}\ \phi\ e^{{B\over{32\pi^2}}
S_0[\hat g ,\phi]+{A\over{32\pi^2}}S_1[\hat g, \phi]-\eta_2\hat S_{R^2}
-\gamma_2\hat S_R-\lambda_2\hat S_{c.c.}}\eqno(4.5)$$
where $\hat S_{R^2}, \hat S_R$, and $ \hat S_{c.c.}$ are marginal
operators of the free theory given by $S_0$ and $S_1$ and will be discussed
below.

The free theory ($\eta_2, \gamma_2, \lambda_2 =0$) of (4.5) has conformal
anomaly $$a=-A+a_M,\hskip 1in b=-B+b_M\eqno(4.6)$$
where $M$ is the operator (3.3). This can be seen as follows: Setting
$\hat g_{\mu\nu}=\tilde g_{\mu\nu}e^{\phi_0}$ we see from (3.8):
$$\eqalign{\int D_{\hat g}\ \phi\ &e^{{B\over{32\pi^2}}
S_0[\hat g ,\phi]+{A\over{32\pi^2}}S_1[\hat g, \phi]}\cr
&=\int D_{\hat g}\ \phi\ e^{\int d^4x {\sqrt{\tilde g}}
\{{B\over{64\pi^2}}[(\phi+\phi_0)\tilde M(\phi+\phi_0)-\phi_0\tilde M
\phi_0] +{A\over{32\pi^2}}[\tilde F(\phi+\phi_0)-\tilde F\phi_0]\}}\cr
&=e^{-{B\over{32\pi^2}}S_0[\tilde g ,\phi_0]-
{A\over{32\pi^2}}S_1[\tilde g, \phi_0]}
\int D_{\hat g}\ \phi\ e^{\int d^4x ({\sqrt{\hat g}}
{B\over{64\pi^2}}\phi\hat M\phi +{A\over{32\pi^2}}\hat F\phi)}
}\eqno(4.7)$$
by shifting $\phi\rightarrow\phi+\phi_0$ and using the fact
that ${\sqrt g}M$ and ${\sqrt g}F$ are conformally invariant.
So $-A,-B$ are the ``classical" contributions\footnote*{Here
and below I will call these contributions also ``anomalies,"
although they actually arise from the fact that the action is classically
not quite conformally invariant.} to (4.6) and $a_M,b_M$ are
the quantum contributions from $M$. Therefore we see from (4.4) that the
ansatz (4.5) is consistent if
$$A=a_O+a_L+a_{{mat}}+a_M,\hskip 1in B=b_O+b_L+b_{{mat}}+b_M.\eqno(4.8)$$
How do we know that $a_M,b_M$ do not depend on the moduli $m_i$?
The only local scale invariant quantity they could depend on
is $\int{\sqrt{\hat g}}\hat F$,
which is a topological invariant in the case of $W_+=0$.

Why does the free part of $S_{{loc}}$ in (4.2) have to be of the
form of the free part of the induced action $S_i$ again? One can plausibly,
though not rigorously, argue as follows:
there are two ways to obtain the right effective action (4.3) via (4.2);
(a), $S_{{loc}}$ is classically conformally invariant
and $a,b$ come purely from the quantum anomaly or
(b), the ``classical"
variation of $S_{{loc}}$ is of the form of the induced action $S_i$.
In case (a) $a$ and $b$ would be just numbers
that will in general not cancel the anomalies as needed
in (4.4) (multiplying $S_{{loc}}$ by a factor would then not change
the conformal anomaly). Only in case (b) there
are parameters like $A,B$ in $a,b$ that can be adjusted to satisfy (4.4). But
the only local free action whose ``classical"
variation is the induced action is the induced action itself.

Let us now turn to the operators $\hat S_{R^2},
 \hat S_R$ and $\hat S_{c.c.}$ in (4.5). The consistency condition
of invariance under rescaling of the background metric
(in particular, the theory is at a renormalization group fixed
point) means that the integrands of
$\hat S_{R^2}$, the ``dressed" Hilbert-Einstein
action $\hat S_R$ and the ``dressed" cosmological constant
$\hat S_{c.c.}$ must be locally scale invariant operators.
Let us try the ansatz
$$\eqalign{\hat S_{c.c}&=\int d^4 x{\sqrt{\hat g}} e^{2\alpha\phi}\cr
\hat S_R&=\int d^4 x{\sqrt{\hat g}} e^{\beta\phi}(\hat\nabla\phi)^2+...\cr
\hat S_{R^2}&=\int d^4 x{\sqrt{\hat g}} (\hat\nabla\phi)^4+...\cr
}\eqno(4.9)$$
with $\alpha, \beta, $ and ``..."
determined so that the integrands of (4.9) are scaling operators of
conformal dimension $4$,
to cancel the $-4$ from ${\sqrt{\hat g}}$.
In the language of string theory, they
are vertex operators of our theory of
noncritical three branes. All of them should be moduli deformations,
if the background metric $\hat g$ is really fictitious.
So far I have verified this
only for $\hat S_{c.c.}$.
The ``..." includes possible corrections of order
$\eta_2,\gamma_2,\lambda_2$ that
may be needed in order to keep the other operators marginal
as we move away from $\eta_2,\gamma_2,\eta_2=0$.
Some calculations with
$\hat S_{R^2}, \hat S_R$ and $\hat S_{c.c.}$
can also be found in [\ma] (however $\alpha=\beta$ there).

To calculate the (classical plus anomalous)
dimension of $e^{2\alpha\phi}$ with action (4.5) at $\eta_2,\gamma_2,
\lambda_2=0$ one may go to conformally flat $\hat
 g_{\mu\nu}=e^{\phi_0}\delta_{\mu\nu}$ where $\hat M=2\hatbox^2$ and $S_1=0$.
Because of the shift $\phi+\phi_0\rightarrow\phi$ in (4.7), the condition
$$\dim(e^{2\alpha\phi})=4 \ \ \hbox{with action}\ \sim\ S_0$$
is equivalent to the condition
$$\dim(e^{2\alpha\phi})=4-4\alpha\ \ \hbox{with action}\ \sim
\ \int d^4x\ \phi\box^2\phi.\eqno(4.10)$$
Due to the quartic propagator, this four dimensional theory is formally
very similar to an ordinary free scalar field theory in two dimensions.
In particular, $:e^{2\alpha\phi}:$ will be a scaling operator.
Its dimension in (4.10) is now purely anomalous.
It is found from the two-point function
$$<e^{2\alpha\phi(x)}e^{-2\alpha\phi(y)}>\sim e^{-4\alpha^2 \Delta(\vert
x-y\vert)}\sim\vert x-y\vert^{-{8\alpha^2}\over B},\eqno(4.11)$$
where the propagator
$$\Delta(r)={2\over B}\log r\ \ \hbox{with}\ \
-{B\over{16\pi^2}}\box^2 \Delta(r)=\delta(r)$$
of the free theory
has been used. Thus, $\dim(e^{2\alpha\phi})={{4\alpha^2}\over B}$, and
(4.10) becomes:
$$4-4\alpha={{4\alpha^2}\over B}
.\eqno(4.12)$$
This determines $\alpha$ once $B$ is known
\footnote*{see the note added at the end of this paper.}
Similarly, $\beta$ in $\hat S_R=\int{\sqrt{\hat g}}e^{\beta\phi}
(\hat\nabla\phi)^2+...$ is determined by requiring $e^{\beta\phi}$ to have
dimension 2:
$$2-2\beta={{\beta^2}\over B}
.\eqno(4.13)$$
$\alpha$ and $\beta$ are independent of the moduli $m_i$,
for the same reason as
$a_M,b_M$ are.

The result for the dimension of the operator $e^{p\phi}$ agrees with the
result of [\ma]. ($Q^2$ of [\ma] is $-2B$)
Let me briefly point out two differences of section 4
with [\ma], where the theory of the conformal factor was studied as
a `minisuperspace' theory
rather than as gravity with a self--duality constraint:
First, I did not use the symmetry argument that was used
in [\ma] to justify (for conformally flat $\hat g$ only)
that $S_{{loc}}$ in (4.2) is again the Liouville action. Second, in [\ma]
$\alpha$ was equal to $\beta$ in (4.9)
with the consequence that the simultaneous presence of
$\hat S_R$ and $\hat S_{c.c.}$ at the fixed point was inconsistent.
This led to a suggestion about the cosmological constant problem,
to a different value for $\alpha$
and will result in a different
value for the analog of the `$c=1$
barrier' of two dimensional gravity.

As in two
dimensions, if $\Phi_i$ is a scaling operator
of the matter theory with conformal dimension $\Delta_i$, the operator
$$O_i \equiv \int d^4 x \sqrt {\hat g} e^{\gamma_i \phi} \Phi_i$$
with $\gamma_i$ determined analoguously to (4.12) by
$$4-2\gamma_i={{\gamma_i^2}\over B}+\Delta_i
\eqno(4.14)$$
is a marginal operator that can be added to the action, at least
infinitesimally.

Provided that truly marginal operators $\hat S_{R^2},\hat S_R$
can also be found, we can now rewrite (3.10) as
$$\eqalign{\int \prod_i dm_i\
&\det (O^\dagger O)^{-{1\over 2}}_{\hat g }\
\det( L^\dagger L)^{1\over 2}_{\hat g}\
\int D_{\hat g} x\ D_{\hat g}\phi\
e^{-S_{mat}[\hat g,x]-S[\hat g, \phi]},\cr
S[\hat g, \phi]&= {-B\over{32\pi^2}}S_0[\hat g ,\phi]+{-A\over{32\pi^2}}
S_1[\hat g, \phi]\cr &+\eta\hat S_{R^2}[\hat g,\phi]+\gamma\hat
S_R[\hat g,\phi]+\lambda\hat S_{c.c.}[\hat g,\phi]
.}\eqno(4.15)$$
The index of $\eta,\gamma,\lambda$ has been dropped.
(4.15) describes free fields plus marginal interactions
in a gravitational instanton background. $A,B$ are given by (4.8)
and $S_0,S_1,\hat S_{R^2},\hat S_R,\hat S_{c.c.}$ by (3.5) and
(4.9).
\vfill\eject

{\bf{5. Scaling}}

As an application of the preceding let us derive scaling laws by studying
the integral over the constant mode of $\phi$, as is done in two
dimensions${}^{{}^{[\dk],}}$
\Ref\sei{N. Seiberg, Prog. Theor. Phys. S 102, 319 (1990)}.
The fixed volume partition
function at the critical point $\eta,\gamma,\lambda\sim 0$ is defined as
$$Z(V) \equiv \int \prod_i dm_i\ \chi(m_i) \int D_{\hat g} \phi
\ e^{{B\over{32\pi^2}}S_0[\hat g,\phi]+{A\over{32\pi^2}}
S_1[\hat g,\phi]}\ \delta(\int \sqrt{\hat g} e^{2\alpha \phi} - V)\eqno(5.1)$$
with $\chi(m_i)$ as in (3.10).
Under the constant shift $\phi\rightarrow \phi+c$ we see from (3.5):
$$\eqalign{ \delta S_0&=c\int d^4x{\sqrt{\hat g}}
\hat G =32\pi^2c\chi\cr
\delta S_1&=c\int d^4x{\sqrt{\hat g}}\hat F =-48\pi^2c\tau,}$$
where the topological invariants $\chi$
and $\tau$ are the Euler characteristic and  signature of the manifold
($W_+=0$ here):
\Ref\ati{E.g., M.F. Atiyah, N.J. Hitchin and I.M. Singer,
Proc. Roy. Soc. London A362,425 (1978)}
$$\tau= {1\over {48\pi^2}}\int d^4 x {\sqrt g} (W_+^2 - W_-^2)
\qquad\hbox{and}\qquad \chi={1\over{32\pi^2}}\int d^4 x {\sqrt g} G.
$$
{}From this it follows that
$$\eqalign{Z(V)&=e^{(-2\alpha+B\chi-{3\over 2}A\tau)c}Z(e^{-2\alpha c}V)\cr
\rightarrow\ \
Z(V) &\sim V^{-1+{1\over{4\alpha}}
(2B\chi-3A\tau)}.
}\eqno(5.2)$$
$\alpha$ is given in terms of $B$ by (4.12).
For the four sphere ($\chi=2,\tau=0$),
$$Z(V)\sim V^{-1+{B\over\alpha}}.$$
This quantity should be the easiest one to check with computer simulations.
\footnote*{See note added at the end of the paper for the value
of the exponent}

Inserting operators into (5.1) yields
$$<O_1...O_n>(V)\sim V^{-1+{1\over{4\alpha}}(2B\chi-3A\tau+2\sum\gamma_i)}$$
with $\gamma_i$ determined by (4.14).
For nonzero cosmological constant $\lambda$ one finds from
$$<O_1...O_n>_\lambda=\int dV e^{-\lambda V}<O_1...O_n>_0(V)\eqno(5.3)$$
the scaling behavior
$$<O_1...O_n>_\lambda\sim\lambda^{-{1\over{4\alpha}}(2B\chi-3A\tau
+2\sum\gamma_i)},\eqno(5.4)$$
provided the integral (5.3) converges, i.e.
${1\over{4\alpha}}(2B\chi-3A\tau+2\sum\gamma_i)>0.$
Otherwise there will be additional cutoff-dependent terms in (5.4).
${}^{[\sei]}$

Replacing in (5.1)
$$\delta(\int \sqrt{\hat g} e^{2\alpha \phi} - V)\ \ \ \rightarrow\ \ \
\delta({{\int \sqrt{\hat g}e^{\beta \phi}[(\nabla\phi)^2+..]}\over
{\int \sqrt{\hat g} e^{2\alpha \phi}}
} - \bar R)$$
one obtains the partition function for fixed curvature per volume at
$\eta,\gamma,\lambda=0$:
$$Z(\bar R)\sim \bar R^{-1+{B\chi\over{\beta-2\alpha}}-{3\over2}
{A\tau\over{\beta-2\alpha}}}.\eqno(5.5)$$

Scaling laws (5.2) and (5.4) are similar to the two dimensional ones.
\vfill\eject

{\bf{6. Outlook}}

{\hskip 1cm \bf{Fixed Points Of Gravity With A Weyl Term}}

Conformally self--dual gravity can also be understood as quantum gravity
with the action
$$\int_{M} d^4 x{\sqrt g}(\lambda +
 \gamma R +\eta R^2 +\rho W_+^2)\eqno(6.1)$$
in the limit $$\rho\rightarrow\infty.\eqno(6.2)$$
(6.1) is the most general local renormalizable fourth order derivative
action of four dimensional gravity, up to the topological invariants
$\tau$ and $\chi$ of the previous section. In section 2, the metric was
split into $\phi$, diffeomorphisms $\xi$, moduli $m_i$ and five $\bar h$
components. (6.2) can be understood as the ``classical limit" for the
$\bar h$ components, in which
only the linearized $W_+$-term
$O^\dagger \bar h$ of (2.5)
is important for the $\bar h$-intgral. This Gaussian integral
can be performed at each point $g_0(\phi,\xi,m_i)$,
$$\int D \bar h e^{-\rho\int d^4x {\sqrt g} W_+^2}\sim
\det(\rho O O^\dagger)_{g_0}^{-{1\over 2}}=
\det(\rho O^\dagger O)_
{\hat g e^\phi}^{-{1\over 2}}.$$
This leads again to the integral (2.6), our starting point. (The extra
factor $\rho$ only renormalizes the cosmological constant and does
not influence the anomaly coefficients of section~3.)

Being fourth order in derivatives,
the $R^2$- and $W_+^2$-terms will give rise to negative norm states.
But this will not bother us in the
limit $\rho\rightarrow\infty$, because the $W_+^2$-term decouples
and we can again tune away the $R^2$-term in the end.
Of course,
the new fourth order derivative action $S_0$ is induced by the conformal
anomaly as in section~3.

One might worry that the renormalization group flow will
take us from $\rho\sim\infty$ to finite $\rho$ so that the limit
(6.2) does not
make sense as an effective theory. However, since at $\rho\sim\infty$
the five $\bar h$ components decouple from the other five components of the
metric, $\rho\sim\infty$ corresponds to a renormalization group fixed
point. More precisely,
defining $\epsilon\equiv{1\over{\sqrt{\rho}}}$,
rescaling $\bar h\rightarrow
\epsilon\bar h$ and expanding the action in $\epsilon$,
one obtains:
$$L_0[\phi,x]+\bar h OO^\dagger\bar h+\epsilon L_i[\bar h,\phi,x]+
o(\epsilon^2)\eqno(6.3)$$
where $L_0$ is the $\bar h$-independent part, and $L_i$ are interaction terms
of $\bar h$ with itself, $\phi$ and $x$, $x$ representing the matter. Thus
the beta function for $\epsilon\sim\rho^{-{1\over2}}$
will receive contributions only from diagrams
that couple $\bar h$ and $\phi$, so it will be at least of order $\epsilon$
and vanish as $\epsilon\rightarrow 0$.
If $(\bar\lambda,\bar\gamma,\bar\eta)$ is a fixed point
of $L_0$, then $(\bar\lambda,\bar\gamma,\bar\eta,\rho
=\infty)$ will be a
fixed point of (6.1).

Presently I do not know whether it will be infrared stable or
unstable.
If it turns out to be infrared stable, (1.1)
describes a ``conformally self-dual phase" of
higher order derivative gravity
-- obviously not the world we live in. If it turns out to be ultraviolet
stable, and if gravity with a Weyl term has something to do with
reality, one could speculate that (1.1) describes gravity at short distances.

Hopefully the results found in section 5
will be the starting point for finding similar results for $\rho$ finite
or zero.

{\hskip 1cm \bf{Topological gravity:}}

In two dimensional quantum gravity the correlation functions of
local operators
are related to the correlation functions of topological gravity
${}^{[\wv]}$ which are intersection numbers of submanifolds
on the moduli space of Riemann surfaces with punctures.
Given the similarity of conformally self-dual quantum gravity
to two dimensional quantum gravity,
it would be very interesting to see if there is a similar relation between
it and four dimensional topological gravity${}^{[\wiw]}$.
This is suggested by the fact that the moduli
space of the latter theory seems to be precisely the moduli space of
conformally self-dual metrics that arose here.
One could look for a matter system, analoguous to the $c=-2$ system in two
dimensions
\Ref\dis{J. Distler, Nucl. Phys. B342, 523 (1990)}
that, coupled to gravity,
reproduces the BRST multiplet of 4D topological gravity. The relation
between the ``topological" and the ``physical"
phase of quantum gravity could then be studied in a less trivial model
than the two dimensional one.

{\hskip 1cm\bf{Conclusion}}

Surprisingly enough, methods of two dimensional
quantum gravity can be applied to four dimensional quantum gravity
at least in the limit of infinite Weyl coupling.
The scaling laws are similar to those of two dimensions
and can hopefully
be compared with numerical simulations based on random triangulations.
\footnote*{predicted values of exponents and analog of $c=1$ barrier
are given in the note below}
It remains to be seen whether analogs of the Virasoro constraints decouple
the negative norm states arising from the fourth order derivative
term that is induced by the conformal anomaly.

Many other interesting questions could now be asked,
but this will be left for the future.

\vskip 1cm

\leftline{\bf{Acknowledgements}}

I would like to thank K. Li, E. Mottola and
M. Staudacher for discussions, J.~H.~Schwarz
for advice and encouragement and E. Witten
for suggesting to study conformally self-dual gravity.
\vskip 1cm

\leftline{\bf{Note added}}

The original version of this paper did not contain the values of
$a_O,b_O,a_L,b_L,a_M$, and $b_M$.
After it appeared, it was pointed out
\Ref\amm{I. Antoniadis, P.O. Mazur and E. Mottola, preprint CPTH-A173.0492
/ LA-UR-92-1483 (1992)}
\Ref\odi{E. Elizalde and S.D. Odintsov, Barcelona preprint UB-ECM-PF 92/12
(1992)}
that the coefficients $a_O+a_L$ and $b_O+b_L$ have been computed long ago
\Ref\ft{E.S. Fradkin and A.A. Tseytlin, Nucl. Phys B 201, 469 (1982)}
and that $a_M,b_M$ for $M$ in (3.3) can also be infered from the literature
\Ref\bvi{A.O. Barvinsky and G.A. Vilkovisky, Phys. Rep. 119, 1 (1985)}
with the results:
$$a_O+a_L={796\over {120}}\hskip 1cm b_O+b_L=-{1566\over{360}}\hskip 1cm
a_M={-8\over {120}}\hskip 1cm b_M=-{-28\over{360}},\eqno(A.1)$$
see [\amm] for independent checks. We can now make some numerical
predictions: with these results one obtains from (3.11) and (4.8)
$$A={1\over {120}}(N_0+6N_{1\over2}+12N_1+788),\hskip .4in
B=-{1\over{360}}(N_0+11N_{1\over2}+62N_1+1538)\eqno(A.2)$$
where $N_0,N_{1\over2},N_1$ are the number of conformally coupled
scalars, spin ${1\over2}$ fermions and massless gauge fields.
(4.12), (4.13) and (4.14) become
$$\eqalign{2\alpha&=-B-\sqrt{B^2+4B},\ \cr
\beta&=-B-\sqrt{B^2+2B}\ ,\ \cr
\gamma_i&=-B-\sqrt{B^2+(4-\Delta_i)B}}\eqno(A.3)$$
Thus $\alpha$ will be real if $B\ge 0$ or $B\le -4$.
The second constraint is the relevant one since $B$ is negative.
The reality constraint $B\le \Delta_i-4$ on $\gamma_i$
is weaker than the one for $\alpha$ in (A.3) as long as we allow
only operators with positive dimension $\Delta_i$.
The signs in front of the square roots have been picked to give the correct
results $\alpha=\beta=1,\gamma_i=2-{\Delta_i\over 2}$
in the classical limit $B\rightarrow -\infty$.

To compare with two dimensional gravity, redefine the anomaly coefficient
$b$ of (3.2) as $\tilde c\equiv -360\ b$, so that
$\tilde c_0,\tilde c_{1\over2},
\tilde c_1$ are 1, 11, and 62. $B\le-4$ becomes
$$\tilde c_{mat}+\tilde c_L+\tilde c_O+\tilde c_M\ge 1440
\rightarrow \tilde c_{mat}\ge -98.\eqno(A.4)$$
The analoguous restriction in two dimensions is $c_{mat}\le 1$, where
$c_{mat}$ is the matter central charge.
If the cosmological
constant term is absent, the barrier for $\tilde c$, rather than being --98,
 is determined by the lowest dimension
operator.
We see that in pure gravity $\alpha$ is
real.
In contrast with two dimensions, the situation is improved by
adding conventional matter like conformally
coupled scalar fields, families of fermions or gauge fields.
The $\tilde c=-98$ barrier would only be crossed by adding exotic matter
with
positive anomaly coefficient $b$ in (3.2). It would be very interesting
to investigate if the barrier becomes positive as we move away from
$\rho\sim\infty$ in (6.1).

Using the values (A.1), we conclude that for conventional matter,
on the sphere and at the critical point, $Z(V)$ always diverges
(faster than $V^{-1}$) at small
volumes and $Z(\bar R)$ at
large curvature per volume. E.g,
for pure gravity on the sphere,
(5.2), (5.5), (A.2) and (A.3) lead to the predictions:
$$Z(V)\sim V^{ -3.675..} \hskip 2cm Z(\bar R)\sim\bar R^{\ +3.194..}
\eqno(A.5)$$

\par\penalty-400\vskip\chapterskip\spacecheck\referenceminspace
   \ifreferenceopen \Closeout\referencewrite \referenceopenfalse \fi
   \line{\fourteenrm\hfil REFERENCES\hfil}\vskip\headskip
   \input referenc.txa
   
\end